\documentclass[twoside]{ilcws10}
\usepackage[latin1]{inputenc}
\usepackage[dvips]{graphicx,epsfig,color}
\usepackage{wrapfig,rotating}
\usepackage{amssymb,amsmath,array}

\begin{document}
\title{
\ Studies on the Electron Reconstruction Efficiency for the Beam Calorimeter of an ILC Detector} 
\author{Olga Novgorodova  on behalf of the FCAL Collaboration
\vspace{.3cm}\\
DESY \\
Zeuthen, Germany
\vspace{.1cm}\\
}

\maketitle

\begin{abstract}
In this talk~\cite{talk_link} recent simulation results on the single high energy electron (sHEe) reconstruction with the Beam Calorimeter (BeamCal) for the ILD detector are presented. Guinea Pig is used to generate the e$^{+}$e$^{-}$ pair background and \textsc{Geant4} for the simulation of electron showers in the calorimeter. An algorithm was developed for the sHEe reconstruction on top of the large e$^{+}$e$^{-}$ background. The efficiency of the sHEe reconstruction is estimated for the nominal and SB-2009 ILC beam parameters.
\end{abstract}

\section{Challenges for the Beam Calorimeter}

For the future International Linear Collider (ILC)~\cite{ILC}, the
Beam Calorimeter is proposed for beam diagnostics and sHEe
reconstruction at low polar angles (5.6 - 45 mrad). In this region
close to the beam pipe, a large number of e$^{+}$e$^{-}$ pairs is
predicted~\cite{talk_link2} due to beamstrahlung photon conversion.
Hence, radiation hard sensors are needed, several materials are under
investigation. SHEe must be detected on top of a widely distributed
pair background. This is a challenge for BeamCal geometry optimization
and the analysis of the signals from this detector.  Such single
electrons or positrons are a product of Standard Model processes,
which form a high background e.g. for SUSY particle
searches. Reconstruction with high efficiency of sHEe in BeamCal has
therefore a crucial importance for ILC physics.
\section{Simulation tools}
\begin{wrapfigure}[14]{r}{0.4\columnwidth}
\vspace{-1\baselineskip}
\includegraphics[width=0.4\columnwidth]{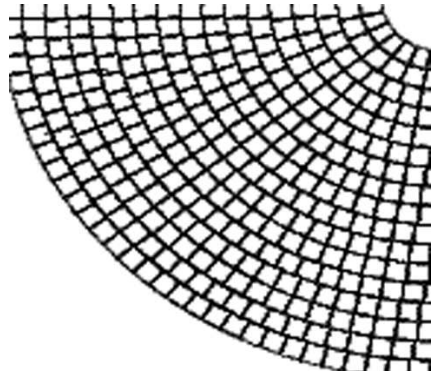}
\caption{ A quarter of the front plane of the Beam Calorimeter. Example of sensor segmentation }\label{Fig:MV}
\end{wrapfigure}

\subsection{Guinea Pig}

For the simulation of the beam-beam interactions at ILC
($\sqrt{s}=500~\textrm{GeV}$), Guinea Pig(GP)~\cite{Guinea-Pig} was
used. Input parameters for GP are the number of particles per bunch,
the energy of the particles, vertical, horizontal and longitudinal
beamsizes, emittances, offsets, etc. The output file ''pairs.dat''
contains e$^{+}$e$^{-}$ pairs from one bunch crossing: the energies,
momenta and initial coordinates of each particle at the interaction
region of the ILC.

\subsection{BeCaS}

For fast simulation and geometry studies, a Beam Calorimeter
simulation tool (BeCaS)~\cite{BeCaS} was written. BeCaS is based on
\textsc{Geant4}~\cite{Geant4} and includes a detailed geometry and
material description of BeamCal as well as a simplified description of
surrounding detectors. A detector solenoid field including an anti-DID
field~\cite{Field}, and a beam crossing angle of 14~mrad are
included. The BeamCal coordinates and dimensions in BeCaS are
$\textrm{z=3550}$~mm, $\textrm{r=20-165}$~mm, and $\textrm{length=120}$~mm. In the BeCaS the
BeamCal is a sandwich calorimeter with 29 layers and 10~cm thick
graphite block on the IP side. An additional layer in front of the
calorimeter plays role of Pair Monitor. Each layer consists of an
absorber, made of 3.5~mm thick tungsten, a sensor plate of 0.3~mm
diamond with gold metalization, a kapton foil of 0.15~mm
thickness and an air gap 0.05~mm thick. The pad size of the sensors is about
$8\times8$~mm$^2$. A sketch of the sensor segmentation is shown in
Figure~\ref{Fig:MV}.

\subsection{Beam parameters}
Two sets of beam parameters are used to generate beamstrahlung pairs:
the ILC nominal beam parameters (NOM)~\cite{Nominal} and the recently
proposed straw-man baseline 2009 beam parameters
(SB-2009)~\cite{SB-2009}.  SB-2009 beam parameters have smaller
horizontal (470~nm) and vertical (5.8~nm) beam sizes than NOM (640~nm and
5.7~nm, respectively). As a result, the beam-beam effect is stronger
and the number of electron-positron pairs in BeamCal is enhanced.

\section{Single High Energy electron reconstruction algorithm}
\begin{wrapfigure}[16]{r}{0.5\columnwidth}
\vspace{-2\baselineskip}
\includegraphics[width=0.5\columnwidth]{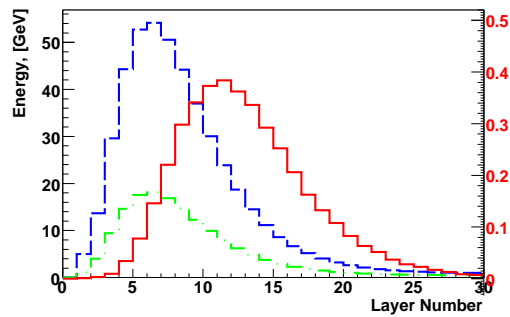}
\caption{Total deposited energy per layer as a function of layer
  number: 250~GeV sHEe (solid line), SB-2009 pair background (dashed
  line), Nominal beam parameters pair background (dash-dotted line)}\label{Fig:Ldistr}
\end{wrapfigure}
To determine sHEe reconstruction efficiency, an algorithm was
developed. First, the average values and the standard deviations of the
deposited energy in every pad of BeamCal are calculated for the pair
background alone, from 10 bunch crossings. The average values of the
deposited energy are then subtracted pad by pad from the deposited
energies obtained from bunch crossings containing sHEe. One standard
deviation of the deposited energy for each pad is applied pad by pad as a threshold for
rejecting number of the opportunities to detect background
fluctuations as a sHEe, in this case called fake electrons. After this
steps, the clustering algorithm is applied.

\subsection{Clustering algorithm}
Electrons and positrons from beamstrahlung pairs have an energy
spectrum around 10~GeV. the maximum deposited energy from beamstrahlung
pairs is located around the 5$^\textrm{th}$ layer of BeamCal for both ILC
beam parameter sets, as shown in the Figure~\ref{Fig:Ldistr}. On the
other hand, the maximum of the sHEe showers is found around the
9$^\textrm{th}$ layer. As mentioned above, and shown in
Figure~\ref{Fig:Ldistr}, in ILC with SB-2009 beam parameters more
beamstrahlung e$^+$e$^-$ pairs are created. Therefore the deposited
background energy in BeamCal for SB-2009 is higher than for nominal ILC beam parameters.
In a first approximation, an sHEe shower is searched for as a cluster
consisting of towers. Every tower must contain 10 or more consecutive
pads with non-zero deposited energy after the 5$^{\rm th}$ layer of
the BeamCal. A cluster candidate must consist of a minimum of two
towers. First, all towers are found, then the tower with the maximum
deposited energy starts a cluster. After this, all neighbouring towers
are searched for and added to the cluster. If a neighbouring tower
has a deposited energy of more than 90\% of the deposited energy of
the initial tower, further neighbouring towers of it are searched for
and added to the cluster. For every cluster, the deposited energy and
the center of mass in radius and polar angle coordinates are calculated.
\subsection{Fake electron subtraction}

\begin{wrapfigure}[31]{r}{0.5\columnwidth}
\vspace{-3\baselineskip}
\includegraphics[width=0.5\columnwidth]{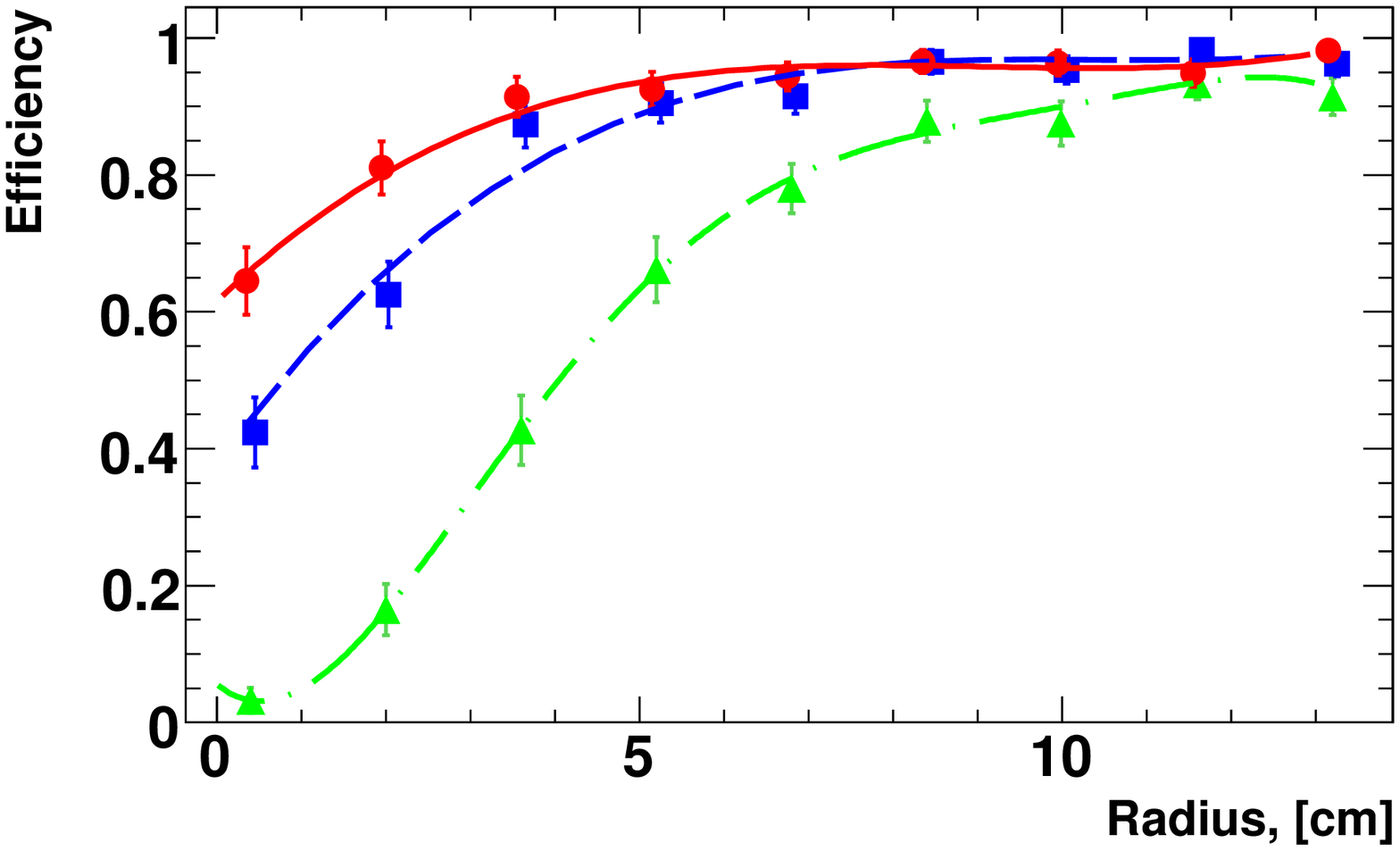}
\caption{Reconstruction efficiency as a function of Radius (as
  measured starting from the outer adge of the beam-pipe) for 50, 150, 250~GeV sHEe for nominal beam parameters}\label{Fig:NOM}
\includegraphics[width=0.5\columnwidth]{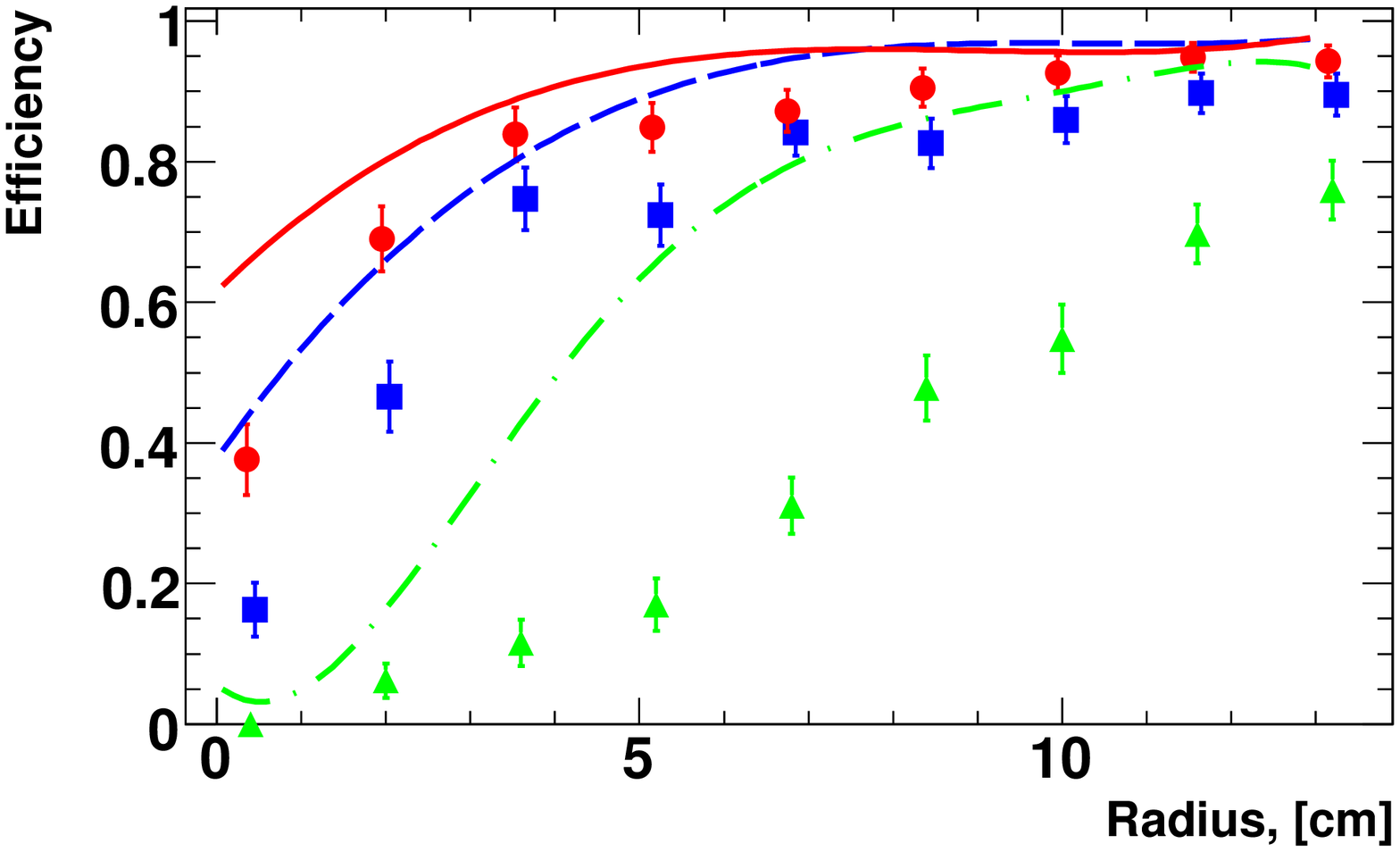}
\caption{Reconstruction efficiency as a function of Radius (as
  measured starting from the outer adge of the beam-pipe) for 50, 150, 250~GeV sHEe for SB-2009 beam parameters}\label{Fig:SB2009}
\end{wrapfigure}

Fluctuations of the background, which are interpreted by the
clustering algorithm as a sHEe, are called fake electrons. By applying
the clustering algorithm to samples of  pure background events the
fake rate was determined. The background distribution decreases with
increasing radius, therefore fake electrons are more likely to appear
at inner radii. Using the method of subtraction of the average value,
and using a threshold of one sigma of the background fluctuations of
signals from every pad, the fake electron rate is reduced. 

\subsection{Reconstruction efficiency}
The reconstruction efficiency is estimated as the ratio between number of reconstructed and generated sHEe,
\[\varepsilon=\dfrac{N_{\mathrm{rec}}}{N_{\mathrm{gen}}} ,\]
Where $N_{\mathrm{rec}}$ - number of reconstructed sHEe, $N_{\mathrm{gen}}$ -
number of generated sHEe. The reconstruction efficiency as a function
of the radial hit position on BeamCal is shown in
Figure~\ref{Fig:NOM}~and~\ref{Fig:SB2009} for nominal and SB-2009 beam
parameter sets. Single high energy electrons of 50~GeV(triangle,
dash-dotted line), 150~GeV(square, dashed line) and 250~GeV(circle,
solid line) are simulated. For every sHEe energy, 1000 files
of electrons were simulated and the clustering algorithm was applied for
each case. A cluster is accepted as a reconstructed sHEe when the
deduced cluster position in radius and polar angle correspond to the
primary position of the sHEe. 
The efficiencies for nominal beam parameters shown in
Figure~\ref{Fig:NOM} are approximated by polynomial functions ( dotted
line curves), which are then overlaid to the result for SB-2009 parameters in Figure~\ref{Fig:SB2009}. The comparison show
that the higher number of e$^{+}$e$^{-}$ pairs created in SB-2009
reduces the reconstruction efficiency considerably, in particular for
lower energies of sHEe. 


\section{Summary and outlook}
A preliminary analysis of single high energy electrons detected in the
ILC BeamCal in the presence of the large background from beamstrahlung
pairs has been presented. So far, the calculations have been performed
with low statistics for the beamstrahlung background. The clustering
algorithm has been developed and applied for nominal and SB-2009 ILC
beam parameter sets for the calculation of the sHEe reconstruction
efficiency.  In comparison with nominal beam parameters, the new
SB-2009 beam parameters show a higher pair background and therefore a
lower reconstruction efficiency. The reconstruction efficiencies of
lower energetic sHEe for SB-2009 drops to less than half the value
obtained for the nominal beam parameters. For higher energetic sHEe,
the reconstruction efficiency is worse by up to 30\%.
This work will be pursued further with adjustment of clustering
algorithm parameters, by adding angular and energy resolutions
calculations, applying the algorithm within the Mokka~\cite{Mokka}
simulations and by using higher statistics.
\section{Acknowledgments}
I would like to thank the Marie Curie Initial Training Network that
has funded this work. I wish to thank the FCAL collaboration, and in
particular Wolfgang Lohmann and Sergey Schuwalow, for valuable discussions.



\begin{footnotesize}


\end{footnotesize}


\end{document}